# Molecular Dynamics Simulation of the Interaction between Cracks in Single-Crystal Aluminum

Hua Ji [a], Keliang Ren [a, *,1], Lihong Ding [a], Ting Wang [a], Jimin Li [a], Jia Yang [a]

[a] *School of Physics and Electronic-Electrical Engineering, Ningxia University, Yinchuan, Ningxia, 750021, China*

**Abstract**

The interaction between cracks, as well as their propagation, in single-crystal aluminum is investigated at the atomic scale using the molecular dynamics method and the modified embedded atom method. The results demonstrated that the crack propagation in aluminum is a quite complex process, which is accompanied by micro-crack growth, merging, stress shielding, dislocation emission, and phase transformation of the crystal structure. The main deformation mechanisms at the front of the fatigue crack are holes, slip bands, and cross-slip bands. During crack propagation, there are interactions between cracks. Such interactions inhibit the phase transition at the crack tip, and also affect the direction and speed of crack propagation. The intensity of the interaction between two cracks decreases with the increase of the distance between them and increases with increasing crack size. Moreover, while this interaction has no effect on the elastic modulus of the material, it affects its strength limit.

## 1. Introduction

Aluminum and its alloys are the most widely used non-ferrous metal structural materials in the industry. Due to its good plasticity, easy processing, and excellent electrical conductivity, thermal conductivity, and corrosion resistance, aluminum has found wide applications in the aviation, aerospace, automotive, machinery manufacturing, shipbuilding, and chemical industries. With the rapid development of modern industrial technology, the working environment of metallic materials and structures has become increasingly harsh, and accidents are often caused by fatigue fracture of components. Therefore, the investigation on the metallic material deformation and fracture processes has important significance to production practice. The fracture of metallic materials is usually affected by internal defects, such as micro-cracks, holes, and dislocations. More specifically, the main cause of the fracture of metallic materials is the instability propagation of micro-cracks. To this end, an increasing number of mechanical and material engineering researchers have begun to investigate and analyze the macro-properties of materials at the micro-scale. The molecular dynamics (MD) method developed in the recent decades provides the function of combining the macro- and nano-characteristics of structures and can offer micro-explanations for several phenomena which are difficult to be explained through theoretical analysis and experimental observation. With the development of computer technology and the related theories, the MD method has become an important research tool of micro/nano-scale fracture mechanics [1]. Nishimura and Miyazaki [2], Inamura et al. [3], Decelis et al. [4], and Swandener [5] performed MD simulations on the crack propagation process of body- or face-centered cubic materials, and explained their crack propagation mechanism. Wu and Yao [6] investigated the effect of temperature on crack propagation of single crystal nickel materials using the MD method. Chandra et al. [7] investigated the crack tip behavior under model I loading and characterized the interaction between the crack tip and vacancies. Uhnakova et al. [8,9] investigated the fatigue behavior of a ductile crack in pure bcc iron under cyclic loading in both modes I and II. They reported that, under uniaxial tensile loads in both modes, the crack could emit dislocations with ⟨1 $\bar{1}$ 1⟩{$\bar{1}$ 1 2} slip system. Cleri et al. [10] performed MD simulations to study the crack-tip shielding by dislocations emitted from the crack tip under monotonic loadings. Sung et al. [11] investigated the crack growth and propagation in pre-cracked single-crystal nickel. Wells [12] reported that crack propagation can be promoted by the coalescence of the main crack and micro-voids. Cui and Beom [13] examined the effects of crack length on the fracture behavior of copper and aluminum single crystals and revealed that the differences in the fracture mechanisms of these two materials were due to their different vacancy-formation and surface energies. According to the above review, MD simulation is an effective method for investigating the behavior of cracks in atomic scale. Moreover, it is of great significance to investigate the fatigue crack propagation mechanism at the micro scale for the anti-fatigue design and engineering application of metal components. While there is a series of studies that, have provided a certain understanding of the initiation and propagation of micro-cracks, only few studies have focused on the interaction of micro-cracks during the expansion process.

---

* Corresponding author
  *Email addresses:* kl_ren@nxu.edu.cn (KL Ren), jihua1202@126.com (Hua Ji)

In order to better understand how micro-cracks effect each other during the propagation process, the propagation process of a multi-crack structure with different crack spacing and size at the atomic scale was simulated and analyzed by the MD method. The simulation results demonstrated that the deformation mechanism around micro-cracks in pure aluminum is a rather complex process, which is accompanied by micro-crack growth, merging, stress shielding, dislocation emission, and phase transformation of the crystal structure. In addition, the distance between cracks and their size were found to affect crack propagation.

## 2. Materials and methods

### 2.1 Molecular dynamics method

The MD method is a simulation method that can solve many-body problems at the atomic and molecular level. It is based on classical Newtonian mechanics, where the interaction between atoms is described by the potential function. In this paper, the modified embedded-atom method (MEAM) potential proposed by Baskes et al. [14,15] is used to describe the interaction between aluminum atoms.

The MEAM potential is a modified form of the embedded-atom method (EAM) suitable for metals and their alloys that have structures such as fcc, bcc, and hcp. In the MEAM theory, the total energy E of the system is expressed as follows:

$$E = \sum_i \left[ F_i(\rho_i) + \frac{1}{2} \sum_{i \neq j} \phi_{ij}(r_{ij}) \right] \quad (1)$$

where $\rho_i$ is the local electronic charge density of the atom $i$, and $F_i(\rho_i)$ is the embedded energy, $\phi_{ij}$ is the two-body potential between atoms $i$ and $j$, and $r_{ij}$ is the distance between atoms $i$ and $j$. The required MEAM parameters of aluminum were obtained from Ref. [15].

In the MD simulation, the position and velocity vectors of an atom at every time step are obtained by integrating the Newton's equation. To this end, Swope et al. [16] proposed the velocity Verlet algorithm based on the work of the Verlet [17], which took the form

$$r_i(t + \delta t) = r(t) + \delta v_i(t) + \frac{1}{2m} F_i(t) \delta t^2 \quad (2)$$

$$r_i(t + \delta t) = v_i(t) + \frac{1}{2m} [F_i(t + \delta t) + F_i(t)] \delta t^2 \quad (3)$$

where $\delta t$ is the integration step. This algorithm can simultaneously store the position, velocity, and acceleration in order to minimize the round-off error.

### 2.2 Definitions of atomic stress and strain

In MD simulations, the definition of stress at the atomic scale differs from that at macroscopic scale. The atomic stress [18], known as the Lagrangian virial stress, is defined as the average of the atomic potentials in the system, which can be described by the following relationship:

$$\sigma_{ij} = \frac{1}{V_i} \left[ \frac{1}{2} \sum_{j \neq i}^{N} r_{ij} \otimes f_{ij} \right] \quad (4)$$

where $f_{ij}$ is the force of atom $j$ acting on atom $i$, $r_{ij}$ is the position vector of atom $i$ relative to atom $j$, and $\otimes$ represents the dyadic operation of two vectors.

The strain under tensile load is described by the following expression [19]:

$$\varepsilon_x = \left| L_{x_0} - L_x \right| / L_{x_0}$$

$$\tag{5}$$

where $L_{x_0}$ and $L_x$ are the respectively initial and total length in the *x* direction.

### 2.3 Molecular dynamics model

Single-crystal aluminum has a face-centered cubic (fcc), structure with a lattice constant of $a_0$=0.405 nm. The dimensions of the MD model were $200a_0 \times 200a_0 \times 5a_0$, as shown in Fig. 1. The total number of atoms in the system was approximately 900,000. The crystal orientations of the *X*-, *Y*- and *Z*-axes were [100], [010], and [001], respectively. The boundary conditions of the model were all periodic boundary conditions. A crack with dimensions of $1.5a_0 \times 4.5a_0 \times 5a_0$ was preinstalled at the center of the model by removing atoms. In order to investigate the effect of the distance between cracks on the failure mechanism, the distance between two cracks varied from $4.5a_0$ to $108a_0$.

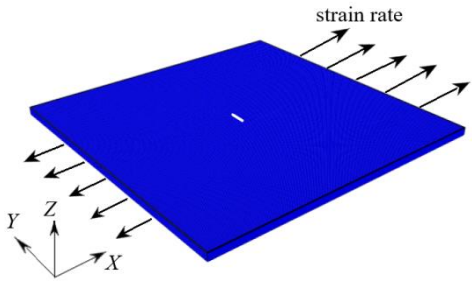
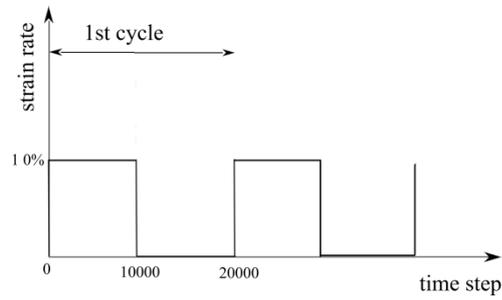

Fig. 1 MD simulation model      Fig. 2 Loading pattern applied to the system

Initially, the velocities of the atoms which follow the Maxwell distribution at a prescribed temperature of 300 K were given. The model was fully equilibrated for 10 ps to stabilize the temperature. During both the equilibration and simulation processes, the time step was 0.001 ps. Subsequently, rectangular cyclic loadings were applied to the system along the crystal direction of [100]. The system was stretched rapidly, while a canonical ensemble was used that maintained the volume and temperature of the system constant. The maximum strain applied to the system in 10,000 time steps was 1%. Then, the strain level returned to 0% and the calculation for relaxation was carried out during 10000 time steps. The same process was repeated 10 times. Fig. 2 presents the loading pattern applied to the system.

### 3. Results and discussion

#### 3.1 Effect of distance between cracks

Fig. 3(b) illustrates the atomic model of 24,000 time steps when the distance between two equally large cracks was $L$=$4.5a_0$. It can be clearly observed that the width of the cracks increased progressively, after the external load was applied. In general, micro-cracks begin to form as the potential between the specific atoms closes. As it can be seen in Fig. 3(a), the preset cracks in the system were rectangular cracks with four sharp angles. Due to stress concentration at the sharp angles, the micro-crack should begin to fracture at the sharp angles. Nevertheless, in the loading process, a stress shield zone was gradually formed between the two cracks, and their propagation inwards was restrained. Consequently, the micro-cracks extended along the [110] and [011] directions. This is consistent with the maximum shear stress direction of isotropic materials. On the other hand, the common neighbor analysis algorithm [20] was employed to analyze the microstructure evolution in the system. As it can be seen in Fig. 3(c), Shockley dislocations were found around the micro-crack after 48,750 time steps through dislocation analysis. In addition, the green and red particles in Fig. 3(c) have fcc and hcp crystal structures, respectively. The gray particles do not have certain crystal structure, and they are concentrated at the regions around the micro-crack tips. Since aluminum is a fcc structure, it can be determined that, in these regions around the crack tips, phase transitions took place. The shape of these phase transition regions is consistent with the results obtained by Nishimura and Miyazaki [2], who suggested that such phase transitions are induced by the high stress which results from the stress concentration around the crack tip. Moreover, Zhao and Chu [21] obtained similar result, and proposed that such phase transitions are related to the atomic potential of the different crystal structures. In fact, stress is concentrated around the crack tip, and disturbs the force balance of the atoms that are located at the fcc crystal structure. As a result, these atoms move, and the crystal structure changes. Furthermore, it was also found that, as the loading continues, both the phase transition and the emission of dislocations play a role in delaying the crack propagation.

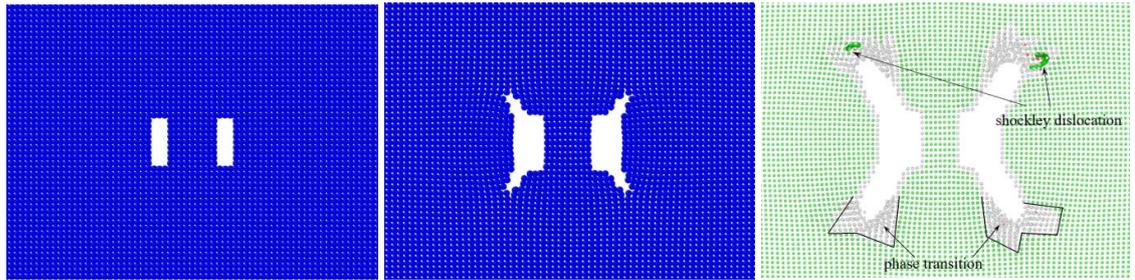

(a) Initial atomic configuration; (b) Atomic configuration at 24,000 time steps; (c) System microstructure and dislocations analysis
Fig. 3 Atomic configuration and system microstructure when $L = 4.5a_0$

As it can be seen in Fig. 4, after 67,500 time steps, a layer of atoms was cut from the (001) crystal plane. In addition, it can be observed that, as the loading continued, the two cracks merged. Voids were formed at micro-crack tip, inducing continuous crack growth. At that point, the crack growth was consistent with that of a single crack. At the same time, slip bands were formed around the crack tip along the [110] and [011] direction, which was the direction of crack propagation. These slip bands, which were consistent with those obtained by Fougeres [22] and Ma et al. [23], should be associated with the maximum shear stress on the 45° plane. This deformation mechanism is called persistent slip bands (PSBs). This consistency also verifies the accuracy of the simulation in this work. It should be noted that, Ma et al. [23] also suggested that the slip bands were able to hinder the initiation and propagation of cracks.

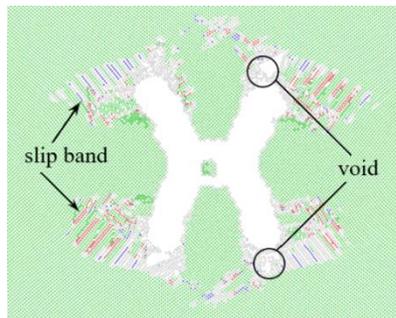

Fig. 4 System microstructure after 67,500 time steps when $L = 4.5a_0$

Fig. 5(a) shows the atomic model, where the distance between two equal-sized cracks was $L=13a_0$, after 25,000 time steps. It can be clearly observed that there was no interaction between the two cracks. Both cracks extended along both the [110] and [011] directions, indicating that with the increase of the distance $L$ between cracks, the interaction between them weakens and the stress shield zone gradually disappears. As it can be seen in Fig. 5(b), after 40,000 time steps, dislocations occurred. (green indicates the Shockley dislocation and magenta the stair-rod dislocation). Moreover, it was also found that the phase change zone was concentrated on the outside of the two cracks. This indicates that with micro-crack growth, there is an interaction between the micro-cracks inside the two cracks, which inhibits the phase transition at the crack tip. At the same time, since there is no interaction at two outer sides of the cracks, the outer micro-cracks expand faster than the inner ones. Fig. 5 (c) shows the microstructure of the system after 62,000 time steps, where it can be seen that as the loading continued, the two inner micro-cracks merged, and the outer micro-cracks continued to expand. Slip bands and cross-slip bands, which are the main deformation mechanism around the crack tip, were also observed.

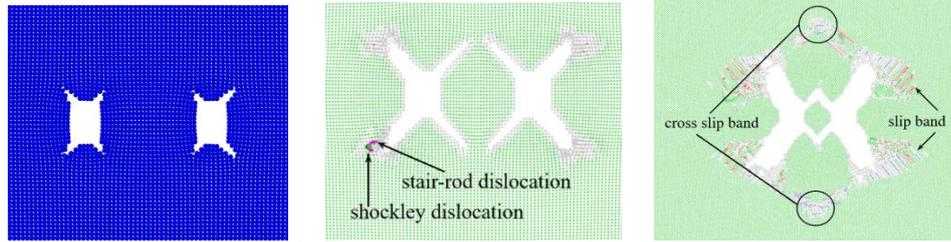

(a) Atomic configurations after 25,000 time steps; (b) System microstructure after 40,000 time steps; (c) System microstructure after 62,000 time steps

Fig. 5 Atomic configurations, dislocation, and the microstructure of the system when $L = 13a_0$

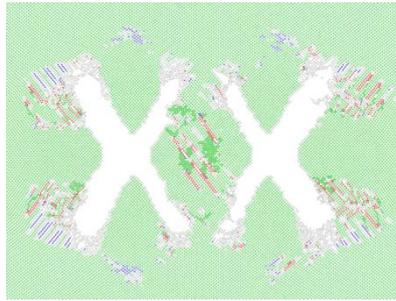

Fig.6 System microstructure after 49,000 time steps when $L = 36a_0$

Fig. 6 shows the microstructure of the system when the distance $L$ between two cracks was $36a_0$. It can be observed that there were also several slip bands in the area between the two cracks. These slip bands were oriented along the [011] direction, which indicated that when two cracks gradually approach each other with the applied load, a stress field is generated between the two cracks. Under the action of stress, slip bands are generated. This also proves that the interaction between cracks is transmitted through the stress field.

Based on the above analysis, it is clear that the intensity of the interaction between cracks depends largely on the distance between them. By performing a series of simulations of systems with the same volume and different crack distances ranging from $4.5a_0$ to $108a_0$, the critical distance for the interaction between two cracks can be determined. When two cracks have the same size and the distance between them is less than $13a_0$, interaction between the cracks occurs; when the distance between them is larger than this critical value, no interaction takes place.

### 3.2 Effect of crack size

In reality, the size of the different cracks in a material is different. To this end, the left-hand crack was defined as I, and the right-hand crack as II, where, the size of crack I was $4.5a_0 \times 1.5a_0$, and that of crack II was 2 and 3 times that of crack I. The distance $L$ between the two cracks varied from $4.5a_0$ to $108a_0$. The simulations were performed under the same cyclic load. The results are shown in Figs. 7 and 9.

According to the simulation results, when the size of crack II was 2 times that of crack I, and the distance $L$ between them was $4.5a_0$, crack II completely inhibited the growth of crack I, and extended along the [110] and [011] directions. As the distance between the two cracks increased, the inhibitory effect of crack II on crack I was gradually weakened. When the distance $L$ between the two cracks was increased to $30a_0$, crack I expanded too; however, crack II expanded faster, indicating that, in this case, the critical distance for the interaction between the two cracks was $30a_0$.

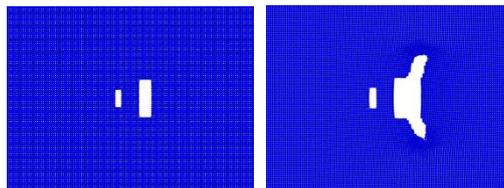

(a) Initial atomic configuration when $L=4.5a_0$; (b) Atomic configuration after 28,000 time steps when $L=4.5a_0$

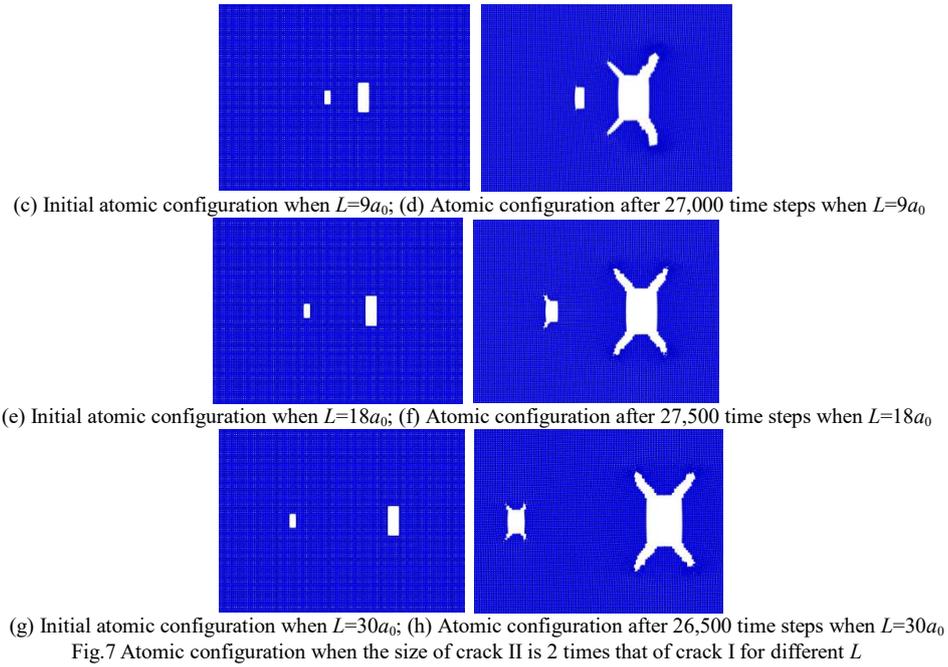

(c) Initial atomic configuration when $L=9a_0$; (d) Atomic configuration after 27,000 time steps when $L=9a_0$

(e) Initial atomic configuration when $L=18a_0$; (f) Atomic configuration after 27,500 time steps when $L=18a_0$

(g) Initial atomic configuration when $L=30a_0$; (h) Atomic configuration after 26,500 time steps when $L=30a_0$
Fig.7 Atomic configuration when the size of crack II is 2 times that of crack I for different $L$

Fig. 8 shows the microstructure of the system when the distance $L$ between the two cracks was $18a_0$ and $36a_0$. As it can be observed, slip bands appeared at the tips of the two cracks. At the same time, the inward propagation direction of crack I changed with $L$. In addition, no slip zone in the area between two cracks appeared, due to the asymmetry of their propagation.

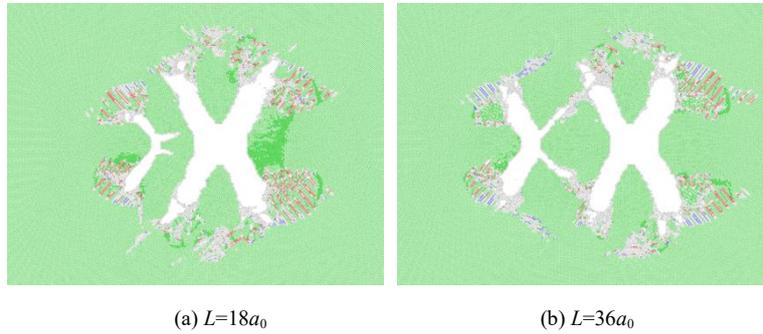

(a) $L=18a_0$          (b) $L=36a_0$

Fig. 8 System microstructure after 48,000 time steps when the size of crack II is 2 times that of crack I for different $L$

When the size of crack II was 3 times that of crack I, and the distance $L$ between them was $18a_0$, crack II completely inhibited the growth of crack I. When the distance between the two cracks was increased to $26a_0$, the interaction between them was weakened, but still existed. As shown in Fig. 9, until the distance $L$ between the two cracks was increased to $36a_0$, there was no interaction between them. This indicates that the critical distance for the interaction between two cracks is $36a_0$. Compared with the results of the previous simulation, it can be clearly deduced that the range of stress shielding area increases with increasing crack size. The results are presented in Fig. 9.

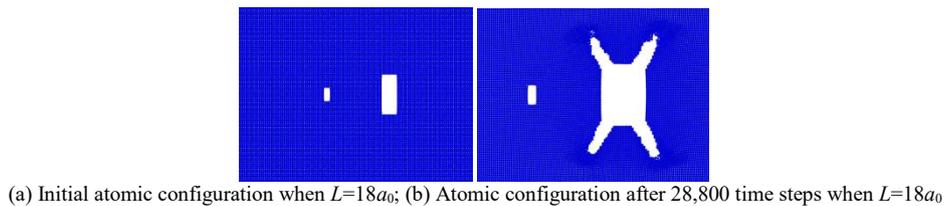

(a) Initial atomic configuration when $L=18a_0$; (b) Atomic configuration after 28,800 time steps when $L=18a_0$

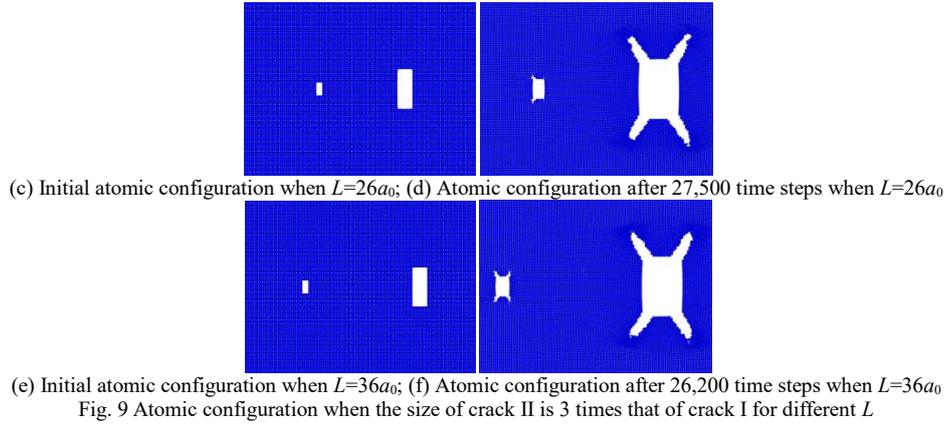
(c) Initial atomic configuration when $L=26a_0$; (d) Atomic configuration after 27,500 time steps when $L=26a_0$

(e) Initial atomic configuration when $L=36a_0$; (f) Atomic configuration after 26,200 time steps when $L=36a_0$
Fig. 9 Atomic configuration when the size of crack II is 3 times that of crack I for different $L$

Based on the above analysis, it can be concluded that the interaction between cracks is not only related to the distance between them, but also to their size. The larger the crack size, the larger the stress shield zone. Moreover, the expansion speed of the larger cracks is significantly accelerated with increasing crack size.

### 3.3 Response to stress-strain

Under room temperature and cyclic loading, microscopic defects such as vacacies, dislocations, and slip bands, are gradually formed in single-crystal aluminum and interact with each other. The formation, and interaction of these microscopic defects affects the mechanical properties of materials, such as plastic deformation, damage, and fracture.

Fig. 10(a) demonstrates the stress-strain relationship of single-crystal aluminum containing one and two cracks of the same size under cyclic loading. Two distances between the two cracks of $4.5a_0$ and $9a_0$ were investigated. As it can be observed, the three curves basically coincided. In particular, the curve of the aluminum with one crack and that with two cracks and $L= 4.5a_0$ completely coincided before failure. This indicates that when the two cracks are very close to each other, they soon merge and expand. Until this point, the interaction between cracks has no effect on the mechanical properties of the material. Fig. 10(b) shows the stress-strain curves of single-crystal aluminum when the distance between the two cracks was, $13a_0$, $36a_0$, $72a_0$, and $108a_0$. It can be observed that the trends of the stress-strain curves of the different models were basically consistent. In particular, the slope of the first half of the stress-strain curve was not affected by the distance between the two cracks, indicating that the micro-defects have no effect on the elastic modulus of the material. However, as the distance between the two cracks increased, the ultimate tensile strength gradually decreased; that is, the ability of the material to resist fracture gradually decreased. Moreover, when the distance between the two cracks was $108a_0$, the stress-strain curve coincided with that when the distance was $72a_0$. This indicated that, when the distance between the two cracks is greater than $72a_0$ and prior to material failure, there is no interaction between the two cracks, and thus, the mechanical properties of the material are not affected.

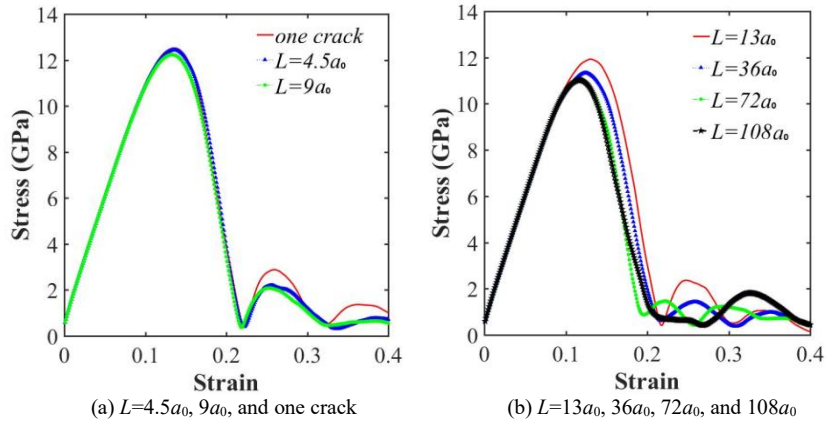
(a) $L=4.5a_0$, $9a_0$, and one crack  (b) $L=13a_0$, $36a_0$, $72a_0$, and $108a_0$
Fig. 10 Stress-strain curves obtained under different $L$

Fig. 11 demonstrates the stress-strain relationships obtained for cracks with different sizes. Figs. 11(a) and (b) show the stress-strain relationship of the material when the distance between the two cracks was $9a_0$a and $72a_0$, respectively. It can be observed that the crack size does not affect the elastic modulus of the material either. At the same time, as the difference in the size of the two cracks increased, the ability of the material to resist destruction gradually deteriorated.

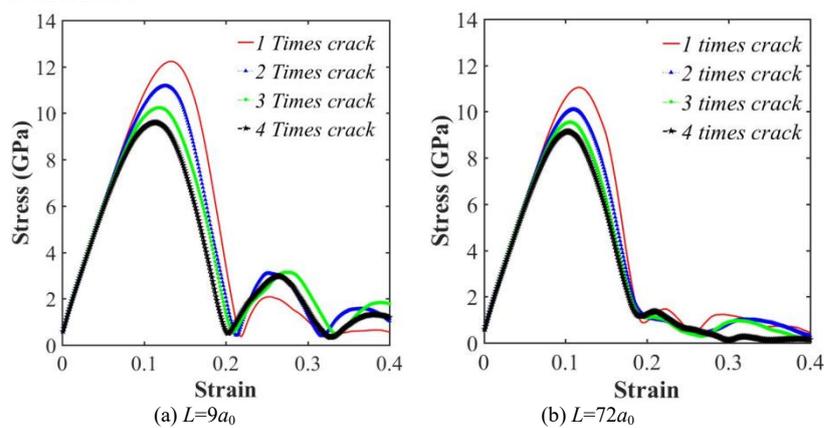

(a) $L=9a_0$      (b) $L=72a_0$
Fig. 11 Stress-strain curves for crack pairs with different sizes

## 4. Conclusions

In this paper, the interaction between cracks, as well as their propagation, in single-crystal aluminum was investigated at the atomic scale using the MD method and MEAM. Considering factors, such as crack size and distance between cracks, the interaction of cracks was determined. Based on the results of the above research, the following conclusions can be drawn:

(i) The crack propagation in aluminum is a quite complex process, which is accompanied by micro-crack growth, merging, stress shielding, dislocation emission, and phase transformation of the crystal structure. Micro-cracks formed around the crack tip, and extended along the [110] and [011] directions. Holes, slip bands, and cross-slip bands were found to be the main deformation mechanisms at the front of the fatigue crack.

(ii) The interaction between cracks is induced through the stress field. This interaction inhibits the phase transition at the crack tip. The intensity of the interaction between cracks depends largely on the distance between them. When the size of two cracks is equal, the critical distance between cracks that enables the interaction between them is less than $13a_0$. When the distance between the cracks is larger than this critical value, no interaction takes place.

(iii) The interaction between cracks is related not only to the distance between them, but also to their size. This interaction affects also the direction and speed of crack propagation. The larger the crack size, the larger the stress shield zone and the stronger the inhibition effect on the smaller crack. Meanwhile, the larger crack propagates faster.

(iv) While the interaction between cracks has no effect on the elastic modulus of the material, it affects its strength limit. The strength limit decreases with the increase of the distance between cracks and the difference in their size.

Finally, it is believed that the proposed work can provide the theoretical basis for anti-fatigue design and process optimization of aluminum alloy parts.

## Acknowledgment

This work was supported by the National Natural Science Foundation of China (No. 11762018) and National key Research and Development Program of China (973 Program, No. 2018YFB2004003).